\documentclass[preprint,aps,prl]{revtex4}
\usepackage{graphicx}

\begin{document}
\title{Novel magnetoinductance effects in Josephson Junction Arrays: A single-plaquette approximation}

\author{Sergei Sergeenkov}

\affiliation{Departamento de F\'isica, CCEN, Universidade Federal
da Para\'iba,\\ Cidade Universit\'aria, 58051-970 Jo\~ao Pessoa,
PB, Brazil}

\date{\today}

\begin{abstract}
Using a single-plaquette approximation, novel magnetoinductance
effects in Josephson junction arrays (JJAs) are predicted,
including the appearance of steps in the temperature behavior of
magnetic susceptibility. The number of steps (as well as their
size) is controlled by the kinetic inductance of the plaquette
whose field dependence is governed by the Abrikosov vortices
penetrating superconducting regions of the array. The experimental
conditions under which the predicted effects should
manifest themselves in artificially prepared JJAs are discussed.\\


\end{abstract}

\maketitle

\section{Introduction}

Many unusual magnetic properties of Josephson junctions (JJs) and
their arrays (JJAs) continue to attract attention of both
theoreticians and experimentalists (for recent reviews on the
subject see, e.g. [1-5] and further references therein). In
particular, among the numerous spectacular phenomena recently
discussed and observed in JJAs we would like to mention the
dynamic temperature reentrance of AC susceptibility [2] (closely
related to paramagnetic Meissner effect [3,4]) and avalanche-like
magnetic field behavior of magnetization [5,6] (well described by
the theory of self-organized criticality [7]). It is also worth
mentioning the recently observed geometric quantization [8] and
flux induced oscillations of heat capacity [9] in artificially
prepared JJAs as well as recently predicted flux driven
temperature oscillations of thermal expansion coefficient [10]
both in JJs and JJAs. At the same time, successful adaptation of
the so-called two-coil mutual-inductance technique to impedance
measurements in JJAs provided a high-precision tool for
investigation of the numerous magnetoinductance (MI) related
effects in Josephson networks [11-14]. To give just a few recent
examples, suffice it to mention the MI measurements [12] on the
JJAs of periodically repeated Sierpinski gaskets which have
clearly demonstrated the appearance of fractal and Euclidean
regimes for non-integer values of the frustration parameter, and
theoretical predictions [13] regarding a field-dependent
correction to the sheet inductance of the proximity JJA with
frozen vortex diffusion. Besides, recently [14] AC
magnetoimpedance measurements performed on proximity-effect
coupled JJA on a dice lattice revealed unconventional behaviour
resulting from the interplay between the frustration $f$ created
by the applied magnetic field and the particular geometry of the
system.

By analyzing the influence of MI on magnetization of SIS-type JJAs
in the mixed Abrikosov state within a single-plaquette
approximation, in the present paper we predict yet another
interesting effect: the appearance of {\it temperature steps} in
the behavior of susceptibility due to vortices induced magnetic
field dependence of the kinetic inductance of the array. We also
discuss the conditions under which the predicted effects can be
observed experimentally in artificially prepared ordered arrays of
unshunted junctions.

\section{The model}

Strictly speaking, to study the influence of MI effects on
temperature behavior of magnetization and susceptibility in
realistic arrays, one would need to analyze in detail the flux
dynamics in these arrays. However, assuming a well-defined
periodic structure of the array (which is actually the case in
most experiments [2,6,8,12,15]), to achieve our goal it is
sufficient to study just a single unit cell (plaquette) of the
array. (It is worth noting that the single-plaquette approximation
has proved successful in treating the temperature reentrance
phenomena of AC susceptibility in ordered 2D-JJA [2,8,15].) Recall
that the unit cell is a loop containing four identical Josephson
junctions. If we apply an external magnetic field $B$ normally to
the plaquette, then the total magnetic flux $\Phi (B)$ threading
the four-junction superconducting loop is given by $\Phi
(B)=BS+L(T,B)I(B)$ where $L(T,B)$ is the loop magnetoinductance
(see below), $S$ the projected area of the loop, and the
circulating current in the loop reads $I(B)=I_C(T)\sin \phi (B)$,
where $\phi (B)=\frac{\pi}{2}[n+\Phi (B)/\Phi _0]$ is the
gauge-invariant superconducting phase difference across the $i$th
junction ($n$ is an integer and by symmetry we assume that [8,15]
$\phi _1=\phi _2=\phi _3=\phi _4\equiv \phi $), $I_C(T)$ is the
field-independent critical current of the junction, and $\Phi _0$
is the magnetic flux quantum.

In turn, the loop magnetoinductance reads $L(T,B)=L_g+L_k(T,B)$
where $L_g=\mu _0a$ is the geometric contribution with $a$ being
the perimeter of the loop, and $L_k(T,B)=\mu _0\lambda ^2(T,B)/a$
is the so-called kinetic contribution with $\lambda (T,B)$ being a
properly defined London penetration depth. In what follows, we
shall assume that the array is in the mixed Abrikosov state  which
means that the field dependence of $L_k$ is due to vortices
penetrating superconducting regions of the array. More
specifically, for $B_{c2}>>B>B_{c1}$, the kinetic
magnetoinductance follows the linear field dependence (dictated by
the corresponding dependence of the penetration depth [16-18])
$L_k(T,B)-L_k(T,0)\propto B$.

As usual, the net magnetization and susceptibility of the plaquette are
given by
\begin{equation}
M(T,B)=-\frac{1}{V}\left (\frac{\partial {\cal H}}{\partial B}\right )
\end{equation}
and
\begin{equation}
\chi (T,B)=\frac{\partial M}{\partial B}
\end{equation}
respectively, where
\begin{equation}
{\cal H}=J(T)[1-\cos \phi (B)]+\frac{\Phi ^2}{2L(T,B)}
\end{equation}
is the Hamiltonian of the system which describes the tunneling
(first term) and magnetoinductive (second term) contributions to
the total energy of the plaquette. Here, $J(T)=(\Phi _0/2\pi
)I_C(T)$ is the Josephson coupling energy, and $V$ the sample's
volume.

\section{Results and Discussion}

To properly address the influence of true MI effects on magnetic
properties of the array, in what follows we assume that for all
applied fields $L(T,B)I(B)\gg BS$. In this approximation, the net
susceptibility of the plaquette will depend on applied magnetic
field only via the universal parameter $\beta _L(T,B)=2\pi
I_C(T)L(T,B)/\Phi _0$, namely
\begin{equation}
\chi (T,B)\simeq -\chi _0(T)\left (\frac{\partial \beta
_L}{\partial f}\right )^2 \left (\cos \beta _L-\frac{1}{2}\beta _L
\right )
\end{equation}
where $\chi _0(T)=I_C(T)S^2/(\pi \Phi _0V)$, and $f=BS/\Phi _0$ is
the frustration parameter.

\begin{figure}
\includegraphics[width=9.5cm]{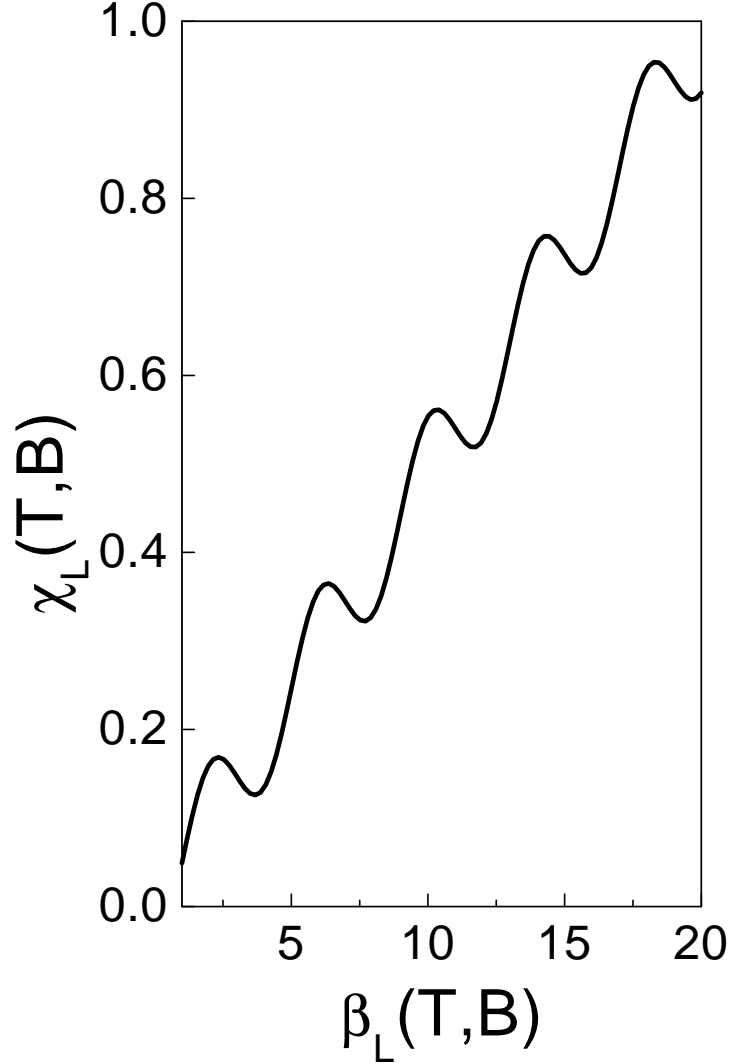}
\caption{\label{fig:1}The normalized magnetoinductance induced
contribution  to susceptibility of a single plaquette $\chi
_L(T,B)$ as a function of the universal parameter $\beta _L(T,B)$,
according to Eq.(4).}
\end{figure}

Let us analyze the obtained results. Figure 1 shows the behavior
of the normalized susceptibility $\chi _L(T,B) \equiv \chi
(T,B)/[\chi _0(T)(\partial \beta _L/\partial f)^2]$ as a function
of $\beta _L(T,B)$ according to Eq.(4). Notice the appearance of
characteristic minima (steps). They correspond to the number of
flux quanta that can be screened by the critical currents in
single plaquette.

To further discuss the predicted behavior of susceptibility, we
need to specify the explicit temperature and field dependencies of
$\beta_L(T,B)$. As usual [8,10], for the explicit temperature
dependence of the Josephson critical current
\begin{equation}
I_C(T)=I_C(0)\left [\frac{\Delta (T)}{\Delta (0)}\right ]\tanh \left [\frac{\Delta
(T)}{2k_BT}\right ]
\end{equation}
we will use the analytical approximation of the gap parameter
(valid for all temperatures) [19], $\Delta (T)=\Delta (0)\tanh
\left (2.2 \sqrt{\frac{T_c-T}{T}}\right )$ with $\Delta
(0)=1.76k_BT_C$.

At the same time, as was mentioned before, the temperature and
field dependencies of the total inductance $L(T,B)$ are governed
by the vortices driven contribution to the kinetic inductance
[16-18], that is $L_k(T,B)= L_k(T,0)[1+B/B_{c1}(T)]$ with
$L_k(T,0)=\mu _0\lambda ^2(T,0)/a=L_k(0,0)/(1-T^2/T_C^2)$ and
$B_{c1}(T)=B_{c1}(0)(1-T^2/T_C^2)$ assuming a two-fluid model
expression for the temperature dependence of the penetration depth
[10].

\begin{figure}
\includegraphics[width=12.0cm]{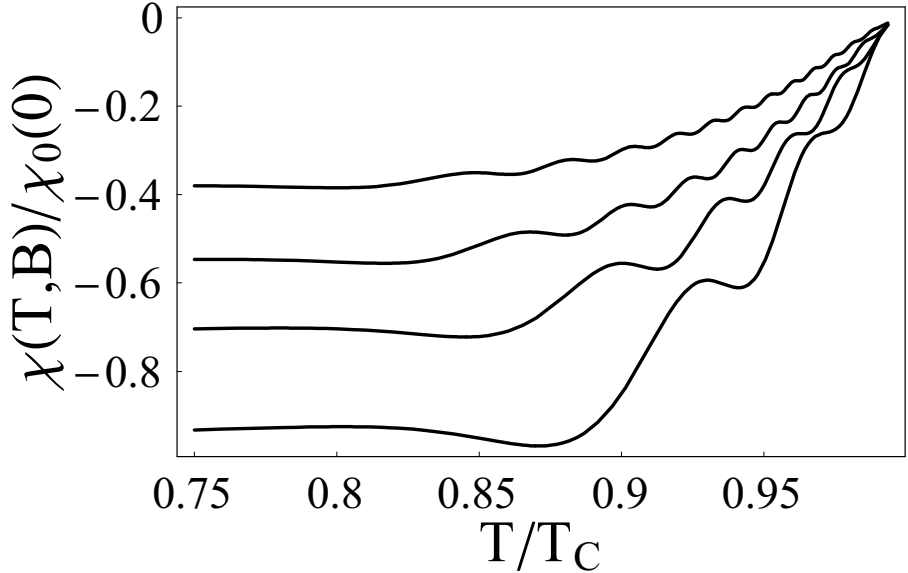}
\caption{\label{fig:2}Theoretically predicted dependence of the
normalized susceptibility on reduced temperature for discrete
values of the universal parameter $\beta _L(0,B)$ (from bottom to
top): $\beta _L(0,B)=\pi ,2\pi ,4\pi $, and $6\pi$.}
\end{figure}

Figure 2 shows the temperature dependence of the normalized MI
induced susceptibility $\chi (T,B)/\chi _0(0)$ for different
values of $\beta _L(0,B)$. Similar to Figure 1, we see the
appearance of well-developed flux-induced {\it temperature steps}
at some discrete values of the critical parameter $\beta _L(0,B)$.
The analysis of Eqs. (4) and (5) near $T_C$ reveals that the
number of steps $n(B)$, their length $\Delta T(B)$ and height
$\Delta \chi (B)$ depend on the value of $\beta _L(0,B)$ as
follows
\begin{equation}
n(B)= \frac{2}{\pi}\beta _L(0,B),
\end{equation}
\begin{equation}
\Delta T(B)= \left [\frac{\pi}{\beta _L(0,B)}\right ]T_C,
\end{equation}
and
\begin{equation}
\Delta \chi (B)= \pi \left [\frac{\partial }{\partial f} \left (
\frac{2\pi}{\sqrt{\beta _L(0,B)}} \right)  \right ]^2\chi _0(0)
\end{equation}
These analytical expressions confirm the predicted correlation
(seen in Fig.2) between increase of the number of steps (as well
as decrease of their size) and magnetoinductance parameter
$L(0,B)\propto \beta _L(0,B)$.

To test experimentally the predicted here effects, two-dimensional
arrays of unshunted $Nb-AlO_x-Nb$ Josephson junctions  with the
following working parameters [8,15] can be used: lattice spacing
$a=46\mu m$ (loop area $S=a^2$), critical current $I_C(4.2K)\simeq
150 \mu A$ for each junction as well as geometric inductance of
the plaquette $L_g=\mu _0a \simeq 64pH$, producing $\beta
_L(0,0)\simeq 30$.

In summary, the influence of magnetoinductance effects on the
temperature behavior of magnetization of the overdamped SIS-type
JJAs in the Abrikosov mixed state was studied theoretically.
Within a single-plaquette approximation, novel phenomenon was
predicted which should manifest itself through appearance of steps
in the temperature behavior of susceptibility. The number of steps
is totally controlled by the vortices driven field-dependent
kinetic inductance of the plaquette.

This work was supported by the Brazilian agency CAPES.

\end{document}